\documentclass[11pt,twoside]{article}
\usepackage{cozumel2005}
\usepackage{epsf}
\usepackage{psfig}
\usepackage{lscape}
\begin{document}
\title{SMC: Stellar Populations through deep CMDs}    
\author{Noelia E. D. No\"el, Carme Gallart}
\affil{Instituto de Astrof\'{\i}sica de Canarias}
\author{Edgardo Costa, Ren\'e A. M\'endez}   
\affil{Departamento de Astronom\'{\i}a, Universidad de Chile}    

\begin{abstract} 
We present deep color-magnitud diagrams (CMDs) reaching the oldest main-sequence turnoffs for 12 fields in the SMC.
The {\it B}-band and {\it R}-band observations were performed using the 100-inch Ir\'en\'ee du Pont telescope at Las Campanas
Observatory, Chile, during four different campaigns (2001-2004). Our fields cover a wide range of galactocentric distance
ranging from $\sim$1$\deg$ to $\sim$4$\deg$  from the center of the galaxy and are located a different position angles. 
Photometry was carried out using DAOPHOT II/ALLSTAR/ALLFRAME. Teramo isochrones have been overlapped.
All our unprecedented deep ground-based CMDs reach the old MS turnoffs with very good photometric accuracy. They clearly show
stellar population gradients as a function of both galactocentric distance and position angle. The most conspicuous difference involves
the young population (age$<$1 Gyr): the young MS is much more populated on the eastern fields, located on the SMC wing area, than on
the western fields located at similar galactocentric radius. In addition, the main stellar population gets progresively older on
average as we go to larger galactocentric radius.  

\end{abstract}


\section{Background}

Dwarf galaxies are believed to represent the dominant population, by number, of
the present day Universe, and a major constituent of groups and clusters.
Studying their star formation and chemical enrichment histories is the key to understand the evolution of galaxies
on cosmological time scales (e.g. Madau {\it et al.} 1998). Local Group galaxies
are ideal laboratories to detailed studies
of dwarf galaxy properties: we can resolve their individual stars and thus learn about their star formation histories (SFH)
by exploring ages, metallicities, and the spatial distribution of the stellar populations they contain. 
The color-magnitude diagram (CMD) is the best tool for retrieving the SFH of a stellar system. CMDs reaching at least
the brightest part of the RGB or much better, the oldest main-sequence (MS) turnoffs, display stars born throughout the
life-time of a galaxy and are fossil records of its SFH. At 50 and 60 kpc, our nearest irregular neighbours, the Magellanic
Clouds (MCs) are an optimal workplace to carry out detailed SFH studies. Thanks to their proximity to the Galaxy, it is possible to
obtain deep CMDs well down the MS turnoff stars (M$_{V}$$\sim$+4). 
 In spite of the proximity and the intrinsec interest, is
remarcable that there are still important gaps in the knowledge of the MCs (e.g. Olszewski {\it et al.} 1996). Maybe their
huge projected size, and the considerably big number of stars to be analized explain the situation. Regarding the field population of the
Small Magellanic Cloud (SMC), this galaxy lacks large area studies reaching faint limiting magnitudes, equivalent to those of  
Gallart {\it et al.} (2004) and  Gallart {\it et al.} (2005) for the Large Magellanic Cloud (LMC).

Gardiner \& Hatzidimitriou (1992) were pioneers in relatively deep wide field studies in the SMC for ages greater than 0.1 Gyr. They
covered the external region ($\geq$2$\deg$ from the center). From their CMDs (which reached the horizontal branch 
 at R$\sim$20 mag) the authors infered an average age of $\sim$10 Gyr and interpreted the existence of a red horizontal branch 
 as an evidence of a 15-16 Gyr population, comprising around 7\% of
the total mass. However this work is mainly limited by the fact that the CMDs do not reach the old MS turnoffs.

 Dolphin {\it et al.} (2001) presented a
combination of HST and ground-based {\it V} and {\it I} images of a SMC field near NGC 121. From their ground-based images of the field,
they infered a peak of star formation between 5 and 8 Gyr ago, with a medium age of 7.5 Gyr, and that 14\%$\pm$5\% of the stars were formed more 
than 11 Gyr ago. However, because of the small field analized these results should not be considered as representative of the whole SMC.

More recently Harris \& Zaritsky (2004), presented the SFH and chemical enrichment history
 of an 4$\deg$$\times$4.$\deg$5 area centered  on the SMC, based on {\it UBVI} photometry from the Magellanic Clouds Photometric Survey
 (Zaritsky {\it et al.} 1997). They found that the SMC formed $\sim$50\% of its total stellar population prior to 8.4 Gyr ago, and that
 there was a quiescent epoch between 3 and 8.4 Gyr ago, followed by more or less continuous star formation starting about 3 Gyr ago
 and extending to the present. 
 
We present here twelve unprecedented deep 
{\it BR} SMC CMDs ranging from $\sim$1$\deg$ to $\sim$4$\deg$ from the center of the SMC and located at different position angles around the 
galaxy. These CMDs clearly show the stellar population gradients in the SMC. Their depth, reaching the oldest MS turnoffs with very good
photometric accuracy, will allow us to obtain detailed SFHs from all of them, and to investigate, therefore, the variation of the SFH
accross the whole SMC field.


\subsection{Observations, Reductions and Photometry}                

\begin{figure}
\plotone{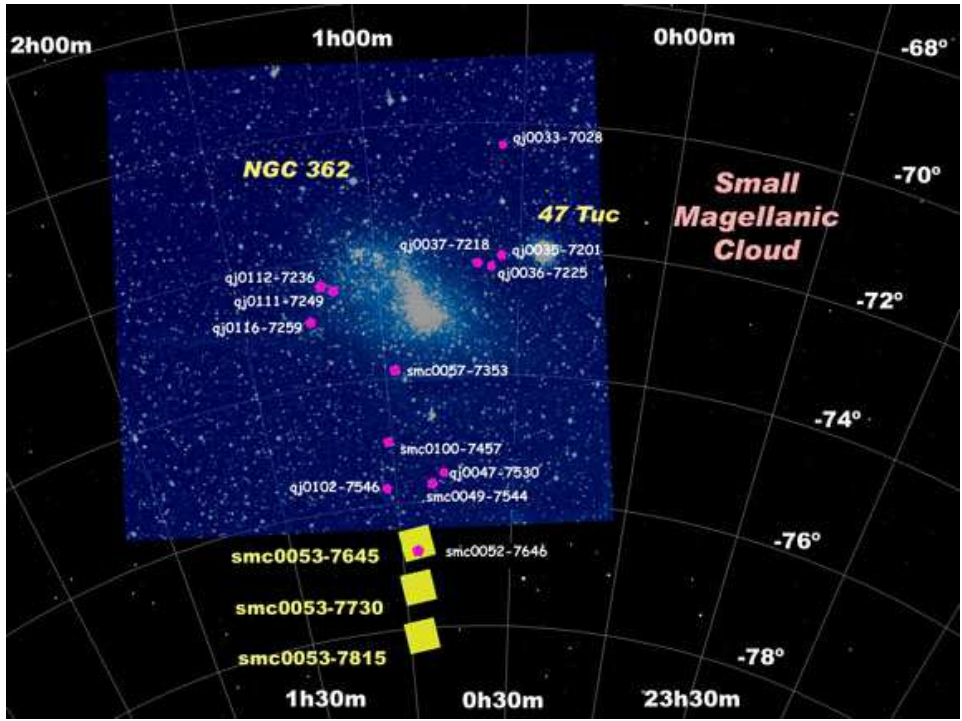}
\caption{Spatial distribution of our SMC fields. The fields obtained during the four campaigns (2001-2004) using LCO 100-inch
are depicted by the hexagons. Big squares denote three fields recently observed using the WFI at the 2.2 m in La 
Silla, Chile. These are considerably larger (34$\arcmin$$\times$33$\arcmin$) to compensate for the lower stellar density
 in the outskirts of the SMC.} \label{campo}
\end{figure}

The observations were made at  
the 100-inch telescope at Las Campanas Observatory. 
Throughout our four years campaign (2001-2004), {\it B}-band and {\it R}-band images of 12 fields in the SMC were obtained.
Each field cover 8.85$\arcmin$$\times$8.85$\arcmin$ 
with a scale of 0.259$\arcsec$/pixel. The fields were chosen to span a wide range of galactocentric distance, from $\sim$1{\deg} to $\sim$4{\deg} 
from the center of the Cloud. \emph{Figure~\ref{campo}} shows the spatial distribution of the 
fields (hexagons).

 In photometric nights, typically six {\it UBVRI} standard star
areas from Landolt (1992) were observed multiple times to determine the
transformation of our instrumental magnitudes to the standard {\it BR} system.  Most of these
areas include stars of a wide range of colors.


Given the crowding in our SMC fields, we performed profile-fitting photometry of the stars in them using the 
 DAOPHOT II/ALLSTAR/ALLFRAME programs, following the steps recommended by Stetson
 (Stetson 1987, Stetson 1990, Stetson 1994). DAOPHOT II/ALLSTAR were used to create a master star list by combining, using
  DAOMASTER (Stetson 1993), the star list obtained for each individual image. Then, ALLFRAME was used to 
 perform simultaneous, consistent reductions of all CCD images of each SMC field. For the final photometry, we selected stars
 with average $\sigma$$\leq$0.15, CHI$\leq$1.8 and -0.3$\leq$SHARP$\leq$0.3.
  We kept a total of 215121 stars, which have been measured in (B-R) down to {\it R}$\leq$25.  
 
\section*{The Color Magnitud Diagrams}         

\begin{figure}
\plotone{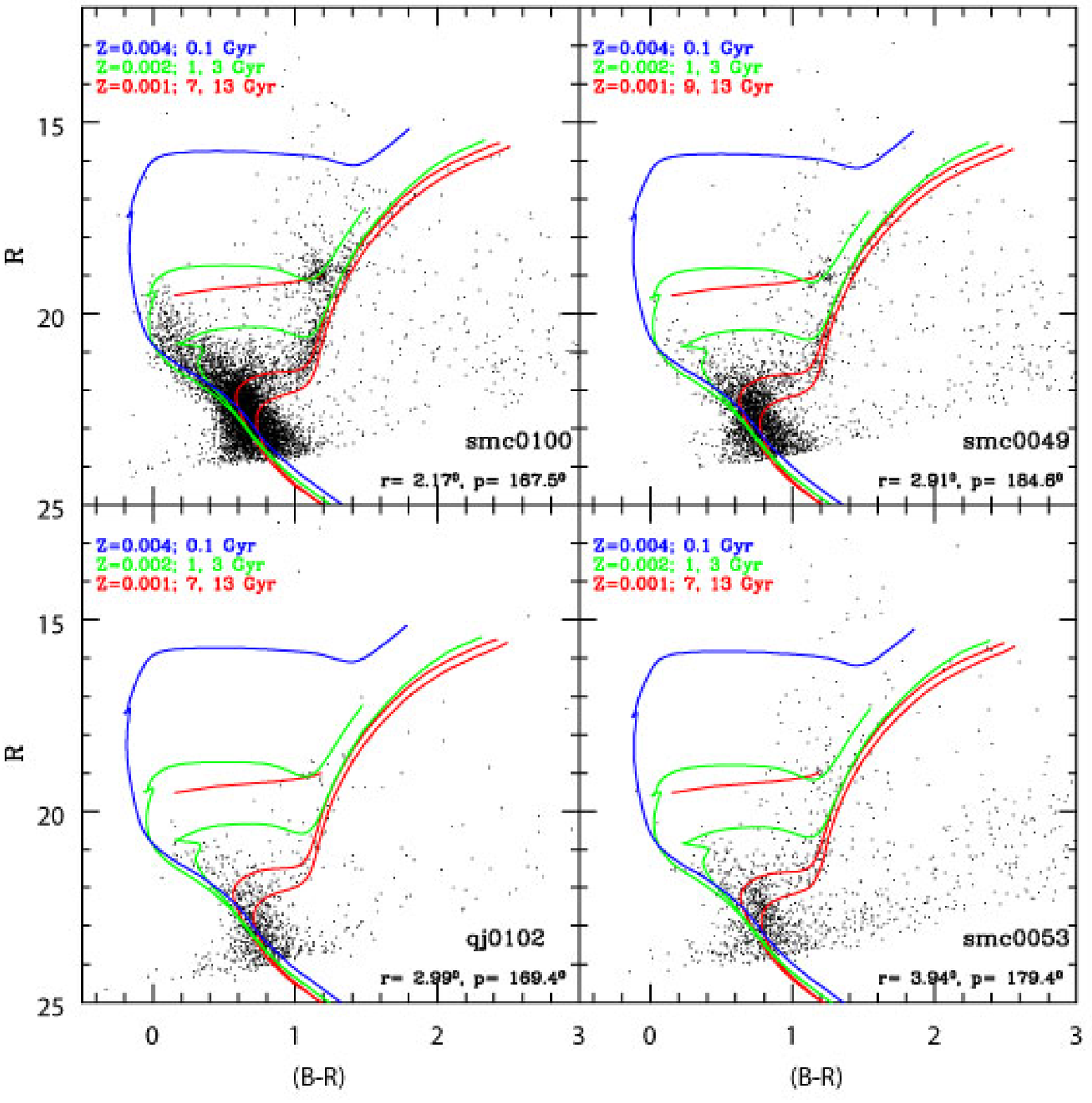}
\caption{SMC CMDs from the southern fields in order of increasing distance. Isochrones from Pietrinferni {\it et al.} (2004) have been
 superposed.} \label{cuatro3}
\end{figure}

\emph{Figures~\ref{cuatro1}} to\emph{~\ref{cuatro3}} show [{\it (B-R), R}] CMDs of the SMC.
A set of Teramo isochrones (Pietrinferni {\it et al.} 2004) have been superposed for three different metallicities
appropiate for stellar populations in the SMC (Z=0.001, Z=0.002,
y Z=0.004). We adopt the distance modulus (m-M)$_{0}$=18.9 (see van den Bergh 1999). 
IRAS/ COBE (Schlegel {\it et al.} 1998) extinction maps
were used to obtain our SMC fields reddening, except for the inner fields smc0057 (at 1.09$^\mathrm{0}$), qj0037 (at 1.31$^\mathrm{0}$),
 qj0036 (at 1.33$^\mathrm{0}$), and qj0111 (at 1.35$^\mathrm{0}$) for which Schlegel {\it et al.} (1998)
 estimated a typical reddening E$_{B-V}$=0.037, from the median
 dust emission in surrounding annuli. Given that their quoted
 value is not accurate, we have estimated a mean value of the reddening 
 for each of the four SMC fields mencioned above.
 In all cases, relations A$_{V}$=3.315E$_{B-V}$,  A$_{B}$=1.316A$_{V}$, and A$_{R}$=0.758A$_{V}$ were used.
 
 \emph{Figure~\ref{cuatro1}} shows SMC fields
qj0112, qj0111, qj0116, and smc0057 corresponding to the eastern side, i.e., facing the LMC, in order of increasing position angle,
 at distances from 1.09$^\mathrm{0}$ to 1.71$^\mathrm{0}$. These diagrams show a prominent, conspicuous MS, which appears populated
  quite  smoothly  from the oldest turnoff at R$\sim$22 to the 0.03 Gyr isochrone. The four CMDs present a substantial number of
   very young stars, 
   which are well matched by isochrones as young as 0.03 and 0.01 Gyr. This very young population may be a consequence of star
  formation triggered by a recent interaction with the LMC and the Milky Way, consistent with the Yoshizawa \& Noguchi (2003) models.

\begin{figure}
\plotone{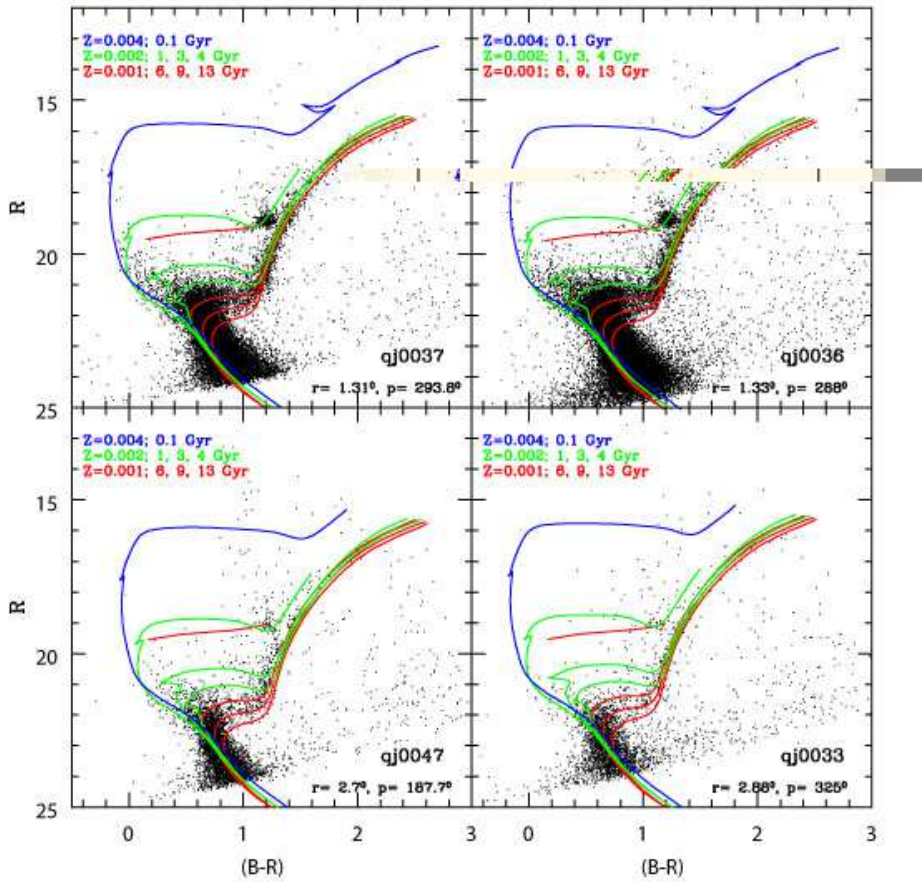}
\caption{SMC CMDs from the western fields. They are ordered at increasing distance from the SMC center.
 The MS appears much less populated than in the eastern side fields. Isochrones from Pietrinferni {\it et al.} (2004) have been
 superposed.} \label{cuatro2}
\end{figure}

\emph{Figure~\ref{cuatro2}} presents four CMDs for fields to the western side of the SMC, 
located up to a distance of $\sim$2.9$\deg$; they are shown in order of increasing distance from the SMC center. 
 It is noticeable that these western fields show a much less populated young MS as compared with the eastern ones at similar galactocentric
 distances. In these fields, the intermediate-age population (1 Gyr$<$age$<$10 Gyr) is the dominant one. Fields qj0036 and qj0037 had star
 formation up to 0.1 Gyr while the others two seem to lack stars younger than 1 Gyr.  

Four CMDs of southern SMC fields, located at distances ranging from $\sim$2.2$\deg$
 to $\sim$4$\deg$, are depicted in \emph{Figure~\ref{cuatro3}}; they are presented in order of increasing distance from the center and
 show a progresively less populated MS in their youngest parts as we go farther south. 
 
 A shared characteristic in all these CMDs is the absense of a populated blue horizontal branch (HB), which is consistent with the fact
 that the position of the $\sim$13 Gyr isochrone is not strongly populated.

\begin{figure}
\plotone{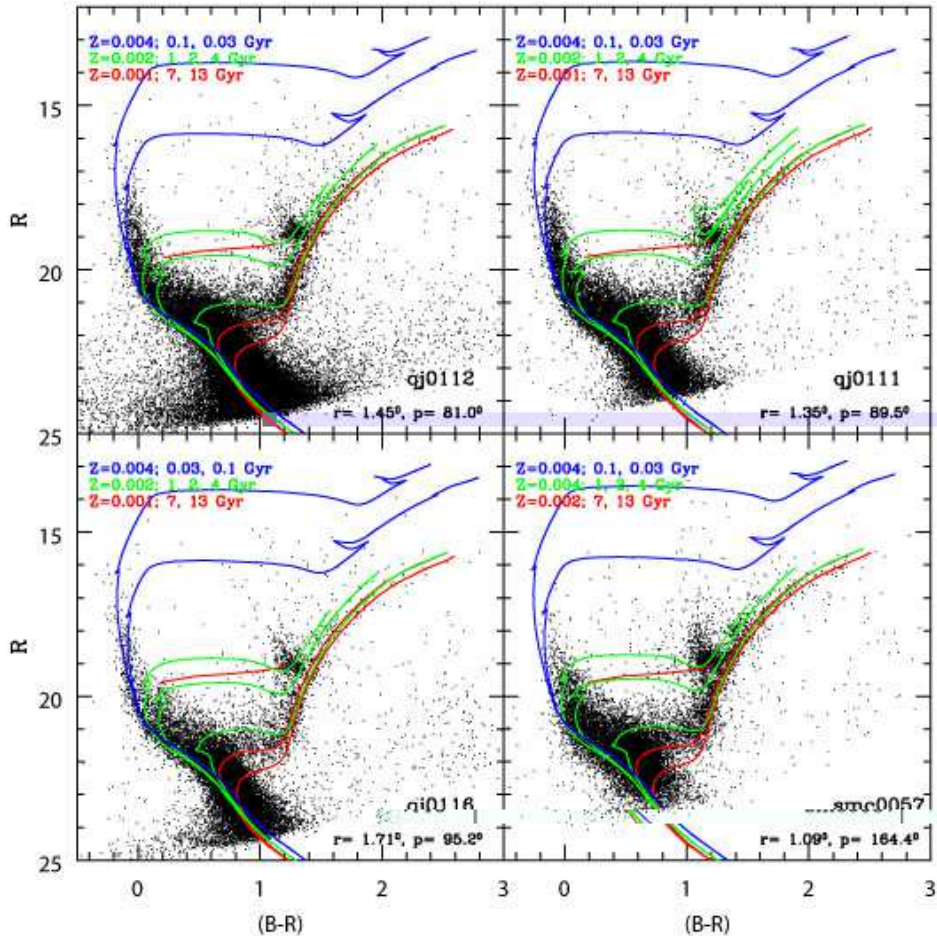}
\caption{SMC CMDs from the eastern fields located at different galactocentric distances at increasing position angles. The conspicuous 
MS which is mostly absent on the western fields, may be the result of star formation triggered by a recent interaction with the LMC 
and the Milky Way. As in the rest of the SMC CMDs there's a lack of HB stars. Isochrones from Pietrinferni {\it et al.} (2004) have been
 superposed.} \label{cuatro1}
\label{cuatro1}
\end{figure}

\subsection*{Discussion}   

These results clearly show the existence of important stellar population gradients as a function not only of the distance to the center
 of the SMC, but also, as a function of the position angle. Thus, a complete description of the stellar populations and evolutive
  history of the SMC requires further analysis involving spatially significant positions. 

\acknowledgements             
The authors acknowledge support by the Plan Nacional de Investigaci\'on Cient\'ifica, Desarrollo, e Investigaci\'on Tecnol\'ogica,
(AYA2004-06343). EC and RAM acknowledge support by the Fondo Nacional de Investigaci\'on Cient\'{\i}fica y Tecnol\'ogica (No. 
1050718, Fondecyt) and by the Chilean Centro de Astrof\'{\i}sica FONDAP (No. 15010003). 


\end{document}